\documentstyle[preprint,aps,eqsecnum]{revtex}
\begin{document}

\title
{Diffusion Thermopower at Even Denominator Fractions}

\author{D.V. Khveshchenko}
\address
{NORDITA, Blegdamsvej 17, Copenhagen DK-2100, Denmark}

\maketitle

\begin{abstract}
\noindent
We compute the electron diffusion thermopower at compressible Quantum Hall states  
corresponding to even denominator fractions in the framework of the composite fermion approach.
It is shown that the deviation from the linear low temperature behavior of the termopower 
is dominated by the logarithmic temperature corrections to the conductivity and not to the
thermoelectric coefficient, although such terms are present in both quantities.
The enhanced magnitude of this effect compared to the zero field case may allow its observation
with the existing experimental techniques.
\end{abstract}
\pagebreak

The thermoelectric effect in metals and semiconductors gives valuable information about the underlying electron-transport processes.
A field of a high current 
activity is the thermoelectric effect in quasi-two-dimensional (2D) 
heterostructure inversion layers.
These systems are characterized by a nearly 2D metallic conduction with very low Fermi energies $E_F\sim 100 K$ 
compared to ordinary metals, which results in a large electron diffusion thermopower. This quantity is experimentally accessible
at low enough temperatures in contrast to most of other parameters related to the thermodynamic properties of 2D electronic
systems, such as the electronic specific heat, which are hardly measurable because of the dominant lattice contribution.

In the presence of an electric field $\vec E$ and the temperature gradient ${\vec \nabla} T$ the electric 
current can be written in terms of the conductivity $\sigma_{ij}$ and the thermoelectric coefficients $\eta_{ij}$ tensors
\begin{equation}
J_i = \sigma_{ij}E_j + \eta_{ij}\nabla_{j} T
\end{equation}
The normally measurable quantity is the thermopower tensor $S_{ij}=-\sigma^{-1}_{ik}\eta_{kj}$ which relates $\vec E$
and  ${\vec \nabla} T$ provided ${\vec J}=0$.

In the presence of a quantizing magnetic field the  $GaAs/Al_{x}Ga_{1-x}As$ heterostructures demonstrate the variety 
of phenomena known as the Quantum Hall Effect. The corresponding behavior of the electron duffusion thermopower 
$S_{ij}$ as a function of the filling factor $\nu ={2\pi n_e\over B}$ is quite
complicated. At $\nu =N+{1\over 2}$ the theory of non-interacting electrons predicts universal peaks of the diagonal termopower
(Seebeck coefficient) $S_{xx}={\log 2\over e(N + 1/2)}$ given by the entropy per particle for the half-filled $N+1^{th}$ Landau level \cite{GJ}
(here and hereafter we assume the carrier charge $e$ to be of either sign). The thermopower of the Integer Quantum Hall states  ($\nu =N$) vanishes at $T=0$, as the entropy
of any number of completely filled Landau levels is zero, and demonstrates the 
thermally activated behavior for small $T$.

In the presence of impurities  the thermopower tensor develops off-diagonal components increasing in magnitude with the strength of disorder for partially occupied Landau levels, which are now broadened into bands. The $S_{xx}$ maxima
at half-integer filling factors, on the contrary,
are predicted to reduce by a factor dependent only on the ratio
between $T$ and the Landau band width \cite{ZL}. It implies, in particular, that the $S_{xx}$ maxima  increase approximately linearly with the magnetic field $B$ as $\nu$ decreases.

The experimental data obtained at $\nu > 4$ agree reasonably well with the above theoretical predicitons.
However, at small $\nu$, where the system is believed to be in the fractional Quantum Hall regime dominated by the electron interactions, both longitudinal and transverse components of the thermopower
behave qualitatively differently. 

In the insulating phases in the vicinity of $\nu =1/3 $ and $2/7$ the diagonal thermopower diverges  at $T\rightarrow 0$
suggesting the spectrum gap \cite{B1}.
One might also expect that the zero entropy argument can be applied to incompressible Quantum Hall states at odd denominator fractions which demonstrate a vanishing diagonal thermopower. At present there are no firm analytical results
available for the FQHE states.

In the present Letter we analyze the behavior of $S_{xx}(T)$ at primary even denominator fractions 
($\nu \sim 1/\Phi$, where $\Phi =2,4$, etc.) which correspond to 
compressible metal-like states. The theory developed by Halperin, Lee, and Read
\cite{HLR} explains the metal-like behavior 
observed at these fractions \cite{FL} by the formation
of the Fermi surface of a new sort of fermionic quasiparticles named composite fermions (CF).

At $T> 0.1K$ the measured thermopower is dominated by phonon drag \cite{Y,B2,T}. 
The phonon drag contribution to $S_{xx}$ scales with temperature approximately as $T^{3.5\pm 0.5}$ and acquires equal values at $\nu$ with the same even denominator, such as
$1/2$ and $3/2$ or 1/4 and 3/4, which agrees well with the theoretical estimate \cite{T}
based on the picture of CF interacting with phonons.

At low enough $T$ the phonon drag term dies off and one expects the diffusion contribution to take over.
To date, the low-temperature measurements which revealed the approximately linear behavior of $S_{xx}(T)$ were only reported on
2D hole systems \cite{Y,B2}. 

The features exhibited by $S_{xx}$ at filling fractions 
$\nu={2N+1\over {4N+2 \pm 2}}$
 in the range from $1/3$ to $2/3$ are strikingly similar to those at high half-integer $\nu$ values \cite{B2}.
This observation lands support to the CF picture where
on the level of the mean field description those fractions correspond to half-filled effective Landau levels 
of CF $(\nu^* =N+{1\over 2})$ in the residual field $B^* =B-2\pi\Phi n_e$ at $\Phi=2$ \cite{HLR}.

The data at the primary fractions
$\nu =1/2$ and $3/2$ were interpreted in \cite{Y} by means of the Mott formula derived for the non-interacting
electrons
at zero field
\begin{equation}
\eta_{xx}=-{\pi^2 T\over {3e}} ({d\sigma (E)\over dE})|_{E=E_F}
\end{equation}
where the energy-dependent Drude conductivity $\sigma (E)= {2\pi e^2\over h}N(E)D(E)$ is a product of the 
density of states $N(E)$ and the diffusion coefficient $D(E)$ 
determined by the transport time $\tau_{tr}(E)\sim E^p$, which results in the classical Drude thermopower
\begin{equation}
S_{xx}={\pi^2 T\over {3e}} ({d\log\sigma (E)\over dE})|_{E=E_F}={\pi^2 (p+1)T\over {3e E_F}}
\end{equation}
In the zero field case the formula 
(3) was conventionally applied to ordinary electrons undergoing scattering from remote as well as
background impurities and from the surface roughness \cite{p}.
Assuming the Matthiessen's rule it was shown in \cite{p} that all  together these three mechanisms lead to a fairly complicated dependence of the exponent $p$ on electron density $n_e$.
At very low densities the scattering by remote ionized donors 
dominates and yields $p=3/2$.
The results obtained at \cite{p} predict that at higher $n_e \sim  1\cdot 10^{11} cm^{-2}$ 
the overall effect of the three
mechanisms can be approximated by $p\approx 1$ while at $n_e \approx 7 \cdot 10^{11} cm^{-2}$ the exponent $p$ 
was found to change sign.

In the framework of the CF theory the only mechanism considered so far was the effect of ionized donors placed on a distance $\xi\sim 10^2 nm$ apart
from the 2D electron gas \cite{HLR}. Under the mapping of electrons at $\nu\sim 1/\Phi$ to the CF the Coulomb impurities become
sources of the spatially random static magnetic field (RMF) correlated as
$<B^*_q B^*_{-q}>= 2\pi^2 \Phi^2 n_i e^{-2\xi q}$, where $n_i$ stands for the impurity concentration.

The RMF scattering appears to be
 essentially more efficient than the ordinary potential one which leads to lower values of $p$. Namely, the result of
 the lowest Born approximation  $\tau^B_{tr}(E)={\xi \sqrt{2E}m^{3/2}\over {\pi\Phi^2 n_i}}$ \cite{HLR}
suggests that at low enough $n_e$ the exponent $p$ might be 
close to 1/2. A more systematic treatment of the RMF problem beyond the
lowest Born approximation \cite{K2} gives 
$\tau_{tr}(E)={2\xi\sqrt{m}\over \sqrt{2E}}\exp({\pi\Phi^2 n_i\over Em})K_{1}({\pi\Phi^2 n_i\over Em})$ which yields an even smaller value $p\approx 0.13$.

In the above discussion of the CF thermopower we neglected the effects of the CF gauge interactions which develop beyond the
mean field approximation \cite{HLR}. 
The available experimental data seem to suggest that the effect of these interactions on $S_{xx}(T)$
can be relatively small in spite of the absence of any small parameter in the present CF theory. Namely, the low-temperature value of the ratio $S_{xx}(3/2)/S_{xx}(1/2)$ was found 
in \cite{Y} to be very close to ${E_F(1/2)\over E_F(3/2)}=\sqrt 3$ in agreement with the mean field picture of the free CF
with the Fermi momentum $k_F=(4\pi n_e(1 -[\nu]/\nu))^{1/2}$ and the effective mass $m^*\sim \nu^{-1/2}$ forming the
metal-like state on the
$[\nu]+1^{th}$ partially occupied Landau level.

Given the complexity of the problem and the previous reports of non-Fermi liquid- type features observed in
resistivity measurements at $\nu =1/2$ and $3/2$ \cite{NFL} this issue deserves further theoretical analysis.

We note at this point that even in the zero field case the effects of
the electron-electron interactions on the 2D thermopower remain poorly understood. Therefore as a prelude to our 
discussion we comment on the 
results obtained in this field so far.

Following the theoretical predictions of the logarithmic temperature corrections to the 2D electrical conductivity
due to both the effects of weak localization and Coulomb interactions  (see, for example, \cite{LR} and references therein)
a similar effect on the thermopower was discussed in \cite{THS}. 
It was argued in \cite{THS} that not only the conductivity $\sigma$ but also the thermoelectric coefficient $\eta$ receive $\log T$ corrections.
This prediction was refuted in a number of subsequent 
publications where it was shown that neither weak localization effects \cite{AGG} nor interference between Coulomb interactions
and impurity scattering \cite{HKR} produce such corrections to the  Peltier coefficient $\Pi$
related to $\eta$ by the Onsager relation
$\Pi =\eta /T$. It was pointed out in \cite{AGG,HKR} that the calculation of  
$\Pi_{ij}=Im {1\over \omega}\int^{\infty}_{0} dt e^{it\omega}<[Q_i(t), J_j(0)]>$ as a 
correlation function of the heat ${\vec Q}={{\vec p}\over m}\epsilon$ and the electric
current ${\vec J}=e{{\vec p}\over m}$ operators requires an application of the finite temperature formalism or an accurate
analytic continuation from imaginary frequencies which had not been done in \cite{THS}.
As a result, the zero field diagonal thermopower was predicted to receive $\log T$ corrections solely
from $\sigma$.
To the best of our knowledge an experimental confirmation  of such terms remains an open question. 

Recently strong although sample-dependent $\log T$ terms in the resistivity at $\nu =1/2$ and $3/2$ were reported \cite{log}.
In the CF picture the impurity scattering is translated to the RMF problem which belongs to the unitary ensemble characterized
by a broken time reversal symmetry. Therefore the localization effects in the RMF are strongly suppressed which rules out
$\log T$ localization corrections \cite{HLR,RMF}. 
On the other hand, it was shown in \cite{K1} that the interference between CF gauge 
interactions and impurity scattering indeed leads to $\log T$ terms which are enhanced by the non-universal factor
as compared to the well-known exchange correction 
$\Delta\sigma_{xx}={e^2\over \pi h}(\log {T\tau_{tr}})$ due to the Coulomb interaction \cite{LR}
which is known to be independent of the magnetic field \cite{HSY}. 

In order to outline the result for $\Delta\sigma^{CF}$ first presented in \cite{K1} and to facilitate a forthcoming calculation 
of $\Delta\eta^{CF}$ we recall that the free CF Green function 
in the presence of 
disorder  $G_{R(A)}(E, {\vec p})={1\over {E -p^2/2m \pm i/2\tau}}$ 
is determined by the (formally divergent) RMF scattering rate
$1/\tau$. It also shows up in the particle-hole diffusion amplitude which develops a pole
(here $\epsilon =E-E_F$)
\begin{equation}
\Gamma (\epsilon, \omega, q)={1\over {2\pi N(E)\tau^2}}{1\over {i\omega - D(E)q^2}}
\end{equation}
at $\omega\tau_{tr}<1$ and  $ql<1$ $(l=v_F\tau_{tr})$ provided that $\epsilon(\epsilon +\omega)<0$ \cite{RMF}.
The gauge invariant physical observables are, however, independent of $\tau$
\cite{RMF}.

In \cite{K1} we showed that 
the main negative temperature correction to the CF magnetoconductivity tensor
$\sigma^{CF}_{ij}$ comes from the transverse vector
coupling mediated by the propagator (here $E$ is the energy of the CF emitting the gauge boson)
\begin{equation}
{\cal D}_{\perp}(\omega, q)={1\over {-i N(E)D(E)\omega + \chi_q q^2}} 
\end{equation}
where $\chi_q={1\over {4\pi^2 N(E)}}({1\over 6}+{1\over \Phi^2})+{1\over (2\pi\Phi)^2}V_q$ is
the CF diamagnetic susceptibility determined by the form of the pairwise electronic potential $V_q$.

In contrast to the zero field case \cite{HSY} this correction depends on the effective magnetic field
$B^*$ according to the relations:
\begin{equation}
\Delta\sigma^{CF}_{xx}(B^*)= (1-(\Omega^*_c\tau_{tr})^2)\Delta\sigma^{CF}_{xx}, ~~~~~~~
\Delta\sigma^{CF}_{xy}(B^*)= 2(\Omega^*_c\tau_{tr}) \Delta\sigma^{CF}_{xx}
\end{equation}
Keeping the explicit dependence on $E_F$ one can write $\Delta\sigma^{CF}_{xx}$
in the short-range case ($V_q \approx V_0 ={2\pi e^2\over \kappa}$, where $\kappa$ is
a screening constant) as
\begin{equation}
\Delta\sigma^{CF}_{xx}={e^2\over 2\pi h}(\log {T\tau_{tr}}) \log [N(E_F)D(E_F)]
\end{equation}
whereas in the case of the unscreened Coulomb potential $(V_q ={2\pi e^2\over q})$ the double-logarithmic terms arise 
\begin{equation}
\Delta\sigma^{CF}_{xx}={e^2\over 2\pi h}(\log {T\tau_{tr}}) 
[\log [N(E_F)D(E_F)]+{1\over 4}\log {T\tau_{tr}}],
\end{equation}
which reduce the correction (8) with respect  to (7) by a factor of $2$ in the range of temperatures
$E_F {1\over (E_F\tau_{tr})^3} <T<1/\tau_{tr}$ ( at lower $T$ the divergency in (8) is cut off). 
Because of the extra logarithm of $N(E_F)D(E_F)$ which equals ${k_Fl\over 4\pi}$ at $B^*=0$ 
the corrections (7-8) are stronger in samples of higher density and/or mobility.

The measured physical magnetoresistivity tensor is related to the CF magnetoconductivity tensor as follows
\begin{equation}
\rho_{ij}=\sigma^{-1}_{ij}=(\sigma^{CF})^{-1}_{ij}+{2h\over e^2}
\pmatrix{ 0 & -1\cr
1 & 0 \cr}
\end{equation} 
while the physical tensor of termoelectric coefficients $\eta_{ij}$ is  simply equal to $\eta^{CF}_{ij}$.

On the basis of the relation (9) we concluded in \cite{K1} that 
in contrast to the case of the Coulomb interacting
 2D electron gas 
the CF Hall conductivity acquires non-zero $\log T$ corrections
which lead to the minima of the diagonal resistivity $\rho_{xx}$ as a function of $\nu$
in the vicinity of the primary even denominator fractions.
The temperature dependence  of $\rho_{xx}$ at these minima exhibits the $\log T$ behavior
whereas the associated Hall resistivity $\rho_{xy}$ shows no such term.

It has to be noted that the available experimental
data \cite{log} seem to demonstrate a stronger dependence of the coefficient in front of the $\log T$
on the metallicity parameter $k_Fl\sim 15-80$.
A better quantitative agreement with the experiment can, presumably, be achieved 
by means of an account of the higher order effects of the gauge interactions. We believe, however, that higher order contributions
do not alter the $\log T$ behavior of $\Delta\sigma^{CF}_{ij}$ which reflects 
the diffusive character of the low-energy long wavelength CF dynamics.

The formula (9) also allows one to understand the well-pronounced symmetric 
V-shaped maxima of $S_{xx}(B^*)$ at $\nu =1/2$ and $3/2$ 
\cite{Y,B2}.
Namely, by using (9) and the classical Drude-like $\eta_{xy}(B^*)={\Omega^*_c\tau_{tr}\over {1+(\Omega^*_c\tau_{tr})^2}}
\eta_{xx}(0)$ with $\eta_{xx}(0)$ given by (2) one obtains 
$S_{xx}(B^*)-S_{xx}(0)= -{2\pi^2(p+1)\over 3}{E_F T\tau^2_{tr}\over \nu^*}$
in the vicinity  ($\Omega^*_c \tau_{tr}<1$) of the primary fractions. 

Now we return to the question of the interaction correction to $\eta$ using the Keldysh technique.
 The calculations similar to those performed in \cite{THS}
for the zero field  case of ordinary electrons yield the leading correction in the form
($f_{\epsilon}$ is the Fermi distribution function)
\begin{equation}
\Delta\eta_{xx}=-{e\tau^2\over 2}\int {d\omega} {d\epsilon} {\epsilon\over T}
 {\partial f_{\epsilon}\over \partial\epsilon}(2f_{\epsilon +\omega}-1)
N^2(E)D^2(E) Im \int {d{\vec q}\over (2\pi)^2} {\cal D}_{\perp}(\omega, q) \Gamma (\epsilon, \omega, q)
\end{equation}
 which comes from small momenta $(ql<1)$ and energy $(\omega \tau <1)$ transfer processes. 
 
It can be readily seen that the expression (10) vanishes unless one expands the result of the 
$\vec q$-integration in odd powers of $\epsilon /E_F$.
This is an example of the general rule
that the diffusion
thermopower must vanish in the limit of a zero Fermi surface curvature when the particle-hole symmetry gets restored.

In the conventional Coulomb case the integral over $\vec q$ is independent of $\epsilon$ and the expression (10)
vanishes \cite{HKR} in contrast to the prediction made
in \cite{THS}, so that the only contribution to 
\begin{equation}
\Delta S_{xx}=S_{xx}(-{\Delta\sigma^{CF}_{xx}\over \sigma^{CF}_{xx}}
+{\Delta\eta_{xx}\over \eta_{xx}})
\end{equation}
comes from $\Delta\sigma_{xx}$ and behaves as $\Delta S_{xx}\sim -{T\over eE^2_F\tau_{tr}}(\log T\tau_{tr})$
 at $T<1/\tau_{tr}$. 

In our case of the transverse gauge interactions the $\epsilon$-integrand in (10) has an extra $\epsilon$-dependence via
the factor $\log [N(E)D(E)]$ (which is the same as as in (7-8)) arising from the integral over $\vec q$.
As a result, we obtain ($p$ is the same as in (3))
\begin{equation}
\Delta\eta_{xx}=-{e T(p+1)\over 12E_F}\log T\tau_{tr}
\end{equation}
It is worthwhile to note that the corrections (7-8) and (12) satisfy the Mott formula (2).
Another example of an approximate validity of the Mott formula 
is provided by the electron-phonon interaction in the regime $ql<1$ \cite{JM}.

Generalizing the method of \cite{HSY} we also obtain that at finite $B^*$ the corrections to the components of the tensor
$\Delta\eta_{ij}(B^*)$ obey the relations analogous to (6).

Despite the fact that $\eta_{ij}$ receives a non-zero contribution (12), the overall correction to the Drude thermopower
(11) is dominated by  $\Delta\sigma^{CF}_{ij}$ given by (7-8)
 and basically behaves as $\Delta S_{xx}\sim -{T\over eE^2_F\tau_{tr}}(\log T\tau_{tr})\log k_Fl$.

The subject of a
 greater controversy is the role of large momenta transfer $(ql>1)$ processes in the thermopower renormalization. 
These effects were previously discussed in both cases of the 
3D electon-phonon \cite{RS} and electron-electron \cite{R} interactions.
It was argued in \cite{RS,R} that the main contribution to $\eta$ comes from processes involving virtual bosons 
(phonons or plasmons). In the problem at hand this kind of corrections is 
associated with the real part of the transverse gauge propagator (which now 
has to be taken in the form ${\cal D}_{\perp}(\omega, q)={1\over {-i\omega k_F/q +\chi_q q^2}}$
accounting for the anomalous skin-effect)
\begin{eqnarray}
\Delta^{'}\eta_{xx}\sim {e\tau_{tr}\over m^4} \int {d\omega} \int {d\epsilon} {\epsilon\over T} 
{\partial f_{\epsilon}\over \partial\epsilon}(2f_{\epsilon +\omega}-1)\nonumber\\
\int {d{\vec q}\over (2\pi)^2}
 Re {\cal D}_{\perp}(\omega, q)
\int {d{\vec p}\over (2\pi)^2}Im G^2_{A}(\epsilon, {\vec p}) Im G_{A}(\epsilon+\omega, {\vec p}+{\vec q})
(p^2+{\vec p}{\vec q})({{\vec p}\times{\vec q}\over q})^2
\end{eqnarray}
as opposed to the conventional kinetic terms 
proportional to $\tau^2_{tr}Im{\cal D}_{\perp}(\omega,q)({\vec p}{\vec q})$ which describe the effects of real
gauge bosons. 
 
It was erroneously stated in a recent publication \cite{SRW} that in the 3D electron-phonon problem
the term similar to (13) (with the factor $\epsilon/T$ absent)
 gives a multiplicative
renormalization of $\sigma_{xx}$ which is essentially  the wave function (or, equivalently, the effective mass)
 renormalization.
In fact, the term discussed in \cite{SRW} 
vanishes in both cases of the electron-phonon and the electron-electron Coulomb interactions in all dimensions \cite{?}.
In the case of the transverse gauge interactions, however, it survives at $T>>E_F/(E_F\tau_{tr})^3$.
And so does the expression (13) which 
then leads to the contribution
$\Delta S_{xx} \sim {T\Phi^2k_F\over e^3E_F}\log {E_F/T}$ for the unscreened Coulomb CF problem 
($\Delta S_{xx} \sim {1\over e}({T\over E_F})^{2/3}$ for the screened case).
Conceivably, this term can be related to the singular CF effective mass renormalization 
($\Delta m^*(T)={\Phi^2k_F\over {2\pi e^2}}\log {E_F/T}$ in the Coulomb case \cite{HLR}) via $E_F\sim 1/m^*(T)$. 

Notably, the Mott formula (2) no longer holds for the above renormalization corrections.

If indeed present (rather than canceled out as a sum of several terms of different origin which all
have a structure similar to (13) \cite{RS,R}), the renormalization correction resulting from large momentum
transfers would appear to be comparable (or even greater)
than the bare Drude
term, which would formally 
invalidate the entire perturbative solution of the quantum Boltzmann equation \cite{RS,R,SRW} based on the assumption
of the dominant impurity scattering.
The work intended to clarify this subtle issue (as well as the effect of interaction corrections to the heat
current operator \cite{HKR}) is now in progress \cite{KR}.

In conclusion, 
we discuss the disorder and interaction effects on the diffusion thermopower of composite fermions in the vicinity of 
the primary even denominator fractions. We show that in contrast to the zero field case of Coulomb interacting electrons
the thermoelectric coefficient of composite fermions
acquires the $\log T$ interference correction resulting from small momentum transfer
processes.
However, the main $T\log T$
correction to the diagonal thermopower arises from the the composite fermion conductivity. The enhanced magnitude of this term compared to the zero field case makes it 
in principle possible to observe such a term experimentally.

\pagebreak

\end{document}